\input phyzzx.tex

\twelvepoint


\def\hg{\hat g}
\def\he{\hat e}
\def\hR{\hat R}
\def\hH{{\hat H}}
\def\hB{{\hat B}}

\def\hF{{\hat F}}
\def\hA{{\hat A}}
\def\uo{{\underline 0}}
\def\uy{{\underline y}}

\def\ub{{\underline b}}
\def\uc{{\underline c}}

\def\un{{\underline n}}


\REF\Polchinski{J. Polchinski, Phys. Rev. Lett. {\bf 75} (1995) 184,
 hep-th/9510017}
\REF\CJS{E. Cremmer, B. Julia and J. Scherk, Phys. Lett. {\bf B76}
(1978) 409}
\REF\Romans{L. Romans, Phys. Lett. {\bf B169} (1986) 374}
\REF\BDPT{E. Bergshoeff, M. de Roo, G. Papadopoulos and P.K. Townsend, 
Nucl. Phys. {\bf B470} (1996) 113, hep-th/9601150}
\REF\BCT{E. Bergshoeff, P.M. Cowdall and P.K. Townsend, {\it Massive IIA
Supergavity from the Topologically Massive D-2-brane}, hep-th/9706094}
\REF\LO{Y. Lozano, {\it Eleven Dimensions from the Massive D-2-brane}, 
hep-th/9707011; T. Ortin, {\it A Note on the D-2-brane of the Massive Type 
IIA Theory and Gauged Sigma Models}, hep-th/9707113}
\REF\Hull{C.M. Hull {\it Gravitational Duality, Branes and Charges}, 
hep-th/9705162}
\REF\CWHMGP{F. Gianni and M. Pernici, Phys. Rev. {\bf D30} (1984) 325;
M. Huq and M. Namazie, Class. Quant. Grav. {\bf 2} (1985) 293;
C. Campbell and P. West, Nucl. Phys. {\bf B243} (1984) 112}
\REF\BDHS{K. Bautier, S. Deser, M. Henneaux and D. Seminara, 
{\it No Cosmological D=11 Supergravity}, hep-th/9704131}
\REF\PT{G. Papadopoulos and P.K. Townsend, Phys. Lett. {\bf B393} (1997)59, 
hep-th/9609095}
\REF\BJO{E. Bergshoeff, B. Janssen and a T. Ortin, {\it Kaluza-Klein
Monopoles and Gauged Sigma Models}, hep-th/9706117}
\REF\CFBH{E. Cremmer and S. Ferrara, Phys. Lett. {\bf B91} (1980) 61;
L. Brink and P.S. Howe, Phys. Lett. {\bf B91} (1980) 384}
\REF\Howe{P.S. Howe, {\it Weyl Superspace}, KCL-TH-96-15}
\REF\GCO{S. Gates, J. Carr and R. Oerter, Phys Lett. {\bf B189} (1987)
68}
\REF\GT{G. Gibbons and P.K. Townsend, Phys. Rev. Lett. {\bf 71} (1993) 3754, 
hep-th/9307049}

\pubnum={KCL-TH-97-46\cr hep-th/9707139}
\date{June 1997}

\titlepage

\title{\bf A New Massive Type IIA Supergravity from Compactification}

\centerline{P.S. Howe\foot{phowe@mth.kcl.ac.uk}${}^{1}$,}
\centerline{N.D. Lambert\foot{lambert@mth.kcl.ac.uk}${}^{1}$}

\centerline{P.C. West\foot{pwest@mth.kcl.ac.uk}${}^{1,2}$}
\address{${}^{1}$Department of Mathematics\break
         King's College, London\break
         England\break
         WC2R 2LS\break
         {}\break
         ${}^{2}$Isaac Newton Insitute,\break
         20 Clarkson Road\break
         Cambridge, England\break
         CB3 OEH}

\abstract

We consider the most general form for eleven dimensional supersymmetry
compatible with on-shell superfields. This allows for the introduction
of a conformal $Spin(1,10)$ connection. 
In eleven dimensional Minkowski space this modification
is trivial and can be removed by a field redefinition, however, upon 
compactification on  $S^1$ it is possible to introduce a non-trivial 
`Wilson line'. 
The resulting ten dimensional supergravity has  massive
1-form and 3-form potentials and a cosmological constant. This theory does
not possess a supersymmetric eightbrane soliton but it does admit a 
supersymmetric non-static cosmological solution.

\endpage


\chapter{Introduction}

Prior to the advent of D-branes as carriers of RR charge [\Polchinski]
the most useful tools in the study of non-perturbative string theory
were supergravities. Indeed even now, next to the newly developed 
and much studied M(atrix) 
theory, eleven dimensional supergravity [\CJS] remains one of the few  tools 
available to study
M theory.

One problem with our present knowledge of M theory is that there is no eleven
dimensional understanding of the type IIA eightbrane.
From the supergravity point of view this can be described  by a 
BPS soliton of Romans'  massive type IIA supergravity [\Romans,\BDPT].
In this case the cosomological constant is interpreted as the Hodge dual
of the eightbrane field strength. There has been some recent work relating
the Romans field equations to the consistency conditions for 
$\kappa$-symmetry of a massive type IIA D-twobrane [\BCT] and the corresponding
picture in eleven dimensions [\LO].
However
there is no known way to obtain the Romans theory directly from eleven
dimensions. Furthermore there has arisen the
possibility of additional type IIA branes [\Hull] for which there is
also no eleven dimensional interpretation. 

Thus it is of interest to understand the origin of massive supergravities by 
compactification from eleven dimensions. In fact since the bosonic
field content of the (massless) type IIA supergravity contains
the $p$-form fields
$(\sigma,A_{m},B_{mn},C_{mnp})$ [\CWHMGP] there
are potentially three distinct Higg's mechanisms which could give rise to 
massess; the vector $A$ could eat
the scalar $\sigma$, the 2-form $B$ could eat the vector, or the
3-form $C$ could eat the 2-form. In Romans theory the 2-form eats the
vector but this leaves open the possibility that  other massive type IIA  
supergravities can be found. Indeed below we shall verify this
and  obtain a new massive type IIA supergravity by compactifying eleven
dimensional supergravity. In this theory the other two Higg's mechanisms are
both employed. In  addition the 
method  we describe appears to be applicable in other dimensions.

Recently it has been argued that there is no possibility for
a cosmological constant in eleven dimensions [\BDHS] which sheads doubt
on the most straightforward interpretation of the type IIA eightbrane as a 
wrapped 
M theory ninebrane (here we mean a BPS soliton of uncompactified eleven 
dimensional supergravity carrying a 10-form charge). Our construction is
consistent with
the conclusions of  [\BDHS]
since the cosomological constant is associated with a compact
direction and so does not appear in eleven dimensional Minkowski space.

The interpretation of the eightbrane as a wrapped ninebrane suffers
from additional problems. In particular the worldsheet effective action 
for an M theory ninebrane would naively be based on a ten dimensional $N=1$ 
supermultiplet
with one scalar. However the only suitable supermultiplets also contain states 
with spins greater than one and have more degrees of freedom than that
contained in the $N=1$, $D=10$ Maxwell supermultiplet which describes the
other D-branes. One the other hand all the D-branes are related by T-duality 
and hence they should all carry the same number of degrees of freedom on their
worldvolumes. A second problem is the lack of
zerobrane/eightbrane type IIA bound states preserving one quarter of 
the spacetime supersymmetry. If the eightbrane is the double dimensional  
reduction of a ninebrane then these states should exist as Kaluza-Klein
modes on the eightbrane [\PT].  Thus one expects that the type IIA 
eightbrane's M theory origin
is like that of the
type IIA sixbrane. Its eleven dimensional interpretation is as a 
Kaluza-Klein monopole which  cannot
exist in the uncompactified  theory. As a result its worldsheet theory
only possesses three scalar fields because the ``zero
mode'' associated to the compact Killing direction is massive in the 
quantum theory (see also [\Hull,\BJO].

For the rest of this paper we shall pursue a construction
of supergravity where the origin of the mass comes from 
topological effects.
We shall see that the massive supergravity we obtain does admit a 
natural BPS state but in contrast to the other supergravities it is
not static. As we mentioned above a cosmological constant is generally 
associated with an eightbrane but here the resulting solution appears to
represent some kind of dynamical instability of the compactification.

\chapter{Weyl Superspace}

In this paper underlined quantities refer to tangent space indices, letters
from the beginning of the alphabet denote eleven dimensional indices and
those from the middle ten dimensional indices. Hatted variables  refer to
eleven dimensional fields and we use the ``mostly plus'' signature. 

In the superspace formulation of eleven dimensional
supergravity the field equations are entirely determined by solving 
constraints on superspace [\CFBH]. The starting point is to introduce a 
spin connection on superspace taking values in $Spin(1,10)$. However it
is  possible to allow for the introduction of a conformal spin
connection [\Howe]. The solution to the constraints then proceeds as before 
provided that the conformal curvature vanishes.

We write the $CSpin(1,10)$ connection as
$$
{\hat \Upsilon}_a = {1\over4}{\hat \Omega}_{a}^{\ \ub\uc}\Gamma_{\ub\uc} 
+ 2 k_a \ .
\eqn\cspin
$$
The condition that the conformal part of the curvature vanishes is then
simply that $dk=0$.
The Lorentz part may be further written  as
$$
{\hat \Omega}_{a\ub}^{\ \ \uc} = {\hat \omega}_{a\ub}^{\ \ \uc} 
+ 2(\he_{a}^{\ \uc}k_{\ub} - \he_{a\ub}k^{\uc})\ ,
\eqn\lspin
$$
where ${\hat \omega}$ is the Levi-Civita connection. The additional terms
in \lspin\ are needed to ensure that the torsion of the connection 
${\hat \Upsilon}$ vanishes. The bosonic degrees of freedom are carried by a 
vielbein $\he_a^{\ \ub}$ and  
anti-symmetric 3-form gauge potential $\hB_{abc}$ with  Weyl weight 2
and field strength $\hH_4 = {\hat D}\hB_3$.
The field equations are [\Howe]
$$\eqalign{
\hR_{ab} - {1\over 2}\hg_{ab}\hR &= -{1\over48}\left(
4\hH_{acde}\hH_b^{\ cde} - {1\over2}g_{ab}\hH^2\right) \ , \cr
{\hat D}^a\hH_{abcd} &= {1\over 36\cdot 48}
\epsilon_{bcde...f...}\hH^{e...}\hH^{f...} \ . \cr}
\eqn\eqone
$$
Note  that 
the curvatures and covariant derivatives appearing in the eleven dimensional
expressions are
those of the conformal spin 
connection ${\hat \Upsilon}$. As it is written equation \eqone\
only considers the bosonic fields. However precisely the same equation is
true if we interpret the fields as being eleven dimensional superfields.
The equations then possess the supersymmetry 
$$\eqalign{
\delta\he_a^{\ \ub}  &= -i\epsilon\Gamma^{\ub}\psi_a\ ,\cr
\delta{\hat B_{abc}} &= -3i\epsilon \Gamma_{[ab}\psi_{c]}\ ,\cr
\delta{\hat \psi}_a  &= {\hat D}_a\epsilon  
+ {1\over36}\left(\epsilon\Gamma^{bcd}{\hat H}_{abcd}
+{1\over8}\epsilon\Gamma_{abcde}{\hat H}^{bcde}\right)\ .\cr}
\eqn\susyone
$$
We note here also the both the massless type IIA supergravity and 
Romans theory can be given a superspace formulation [\GCO].

\chapter{Massive Electrodynamics and Topology}

Before discussing the compactification of eleven dimensional supergravity 
with the addition conformal connection it is instructive to first consider 
the analogous case for electrodynamics. To this end suppose we have the action
$$
S = {1\over 4}\int_{\cal M} d^dx\sqrt{-\hg}{\hat F}_{MN}{\hat F}^{MN} \ ,
\eqn\meta
$$
defined over a $d$ dimensional manifold ${\cal M}$ with a metric $g_{MN}$ 
which we
will assume to be non-dynamical and 
$M,N=0,\ldots,d-1$.  In standard electrodynamics one introduces 
the exterior derivative $d$ and defines $\hF = d\hA$ as the 2-form field 
strength of a
$U(1)$ connection $A$.
However, if we introduce a
1-form $k$ then we could also define another derivative $D = d + k$ and
curvature $\hF=D\hA$. By
taking  $k$ to be 
closed we maintain  $D^2 =0$ and as a result the 
action \meta\ is invariant under the transformation
$$
\hA_M \rightarrow \hA_M +  D_M\lambda \ .
\eqn\trans
$$
Now $\hF$ is by definition 
a closed 
2-form (with respect to $D$) on ${\cal M}$ and the field equation of \meta\ 
states that $\hF$ is also
co-closed, so that if $\cal M$ is compact the classical solutions 
correspond to 
elements of the  cohomology $H^2_D({\cal M})$ of $D$. 

If $k=md\psi$ is exact (for example if ${\cal M}$ is simply connected), then 
we may write $D= e^{-m\psi}\cdot d\cdot e^{m\psi}$ and it follows that
$H_D^*({\cal M})$ is isomorphic to the De Rham cohomology group $H_{DR}^*({\cal 
M})$. 
Furthermore by simply redefining
$\hA_M\rightarrow e^{-m\psi}\hA_M$ (i.e. $A$ has Weyl weight one), 
$\hg_{MN}\rightarrow e^{-{4m\over(d-4)}\psi}\hg_{MN}$ 
we would obtain the action for standard 
electrodynamics (on a comformally equivalent manifold). 

Let us now suppose that ${\cal M}$ is not simply connected and that $k$ is not
exact. In this case the above field redefinition is no longer globally
valid. In particular let us choose coordinates in which 
$k_y = m,\ k_{\mu}=0$ with $\mu=0,\ldots,d-2$  and compactify the theory along
$y$. As is usual we define $A_{\mu} = \hA_{\mu}$ and $\phi = \hA_y$ and take 
them to be
independent of $y$. The equations
of motion are now (we assume here that $\cal M$ is flat for simplicity)
$$\eqalign{
\partial^{\mu}F_{\mu\nu}  - m\partial_{\nu}\phi + m^2A_{\nu} &= 0 \ ,\cr
\partial^2\phi - m \partial^{\mu}A_{\mu} &=0 \ .\cr}
\eqn\eqmo
$$
It is easy to see that the second equation in \eqmo\ is the integrability 
condition for
the first. This  apparent loss of a degree of freedom arises because the
symmetry \trans\ reduces to
$$
A_{\mu} \rightarrow A_{\mu} + \partial_{\mu}\lambda\ , \ \ \ \ \ \ \ \ \ \  
\phi \rightarrow \phi + m\lambda \ .
\eqn\transtwo
$$
Thus we may gauge away $\phi$ and arrive at the equations of massive 
electrodynamics
$$
\partial^2 A_{\mu} + m^2 A_{\mu}=0 \ .
\eqn\eqfinal
$$
Note however that the photon field is tachyonic, indicating that this 
compactification
is somehow unstable.

\chapter{Dimensional Reduction to Ten Dimensions}

Let us now return to the dimensional reduction of eleven dimensional 
supergravity [\CWHMGP].
As in the previous section if $k$ is exact, ie $k=md\psi$, then it 
is possible to redefine the fields
$\he_{a}^{\ \ub},{\hat \psi}_a,{\hat B}_3$ so as to absorb $\psi$ and we arrive
at the standard equations of eleven dimensional supergravity.
Let us now 
consider the case that the eleven dimensional space has the topology 
${\bf M}^{10}\times S^1$. If $y$ is a coordinate for the $S^1$ then we may
let the conformal part of the $CSpin(1,10)$ connection take the form
$$
k = mdy \ .
\eqn\kdef
$$
To compactify we will simply make the usual ansatz that the fields are
independent of $y$. For the purposes of this paper we shall ignore the
fermions in ten dimensions. They may be obtained from the eleven dimensional
superspace formulation and are guaranteed to provide a supersymmetric
completion of the theory. In addition we will  
only consider compactifications with $\hH_{abcd}=0$.
If we make the standard string-frame reduction
$$
\he_a^{\ \ub} = \left(\matrix{
e^{-\sigma/3}e_m^{\ \un} & A_me^{2\sigma/3} \cr
0 & e^{2\sigma/3}\cr}\right)\ , 
\eqn\reduction
$$
we find the following equations of motion for the bosonic fields
$$\eqalign{
R_{mn} - {1\over2}g_{mn}R =
&-{1\over2}\left(F_{mp}F_n^{\ p}-{1\over4}g_{mn}F^2\right) \cr 
&+2\left(D_mD_n\sigma - g_{mn}D^2\sigma + g_{mn}(D\sigma)^2\right) \cr
&+18m\left(D_{(m}A_{n)}-g_{mn}D^pA_p\right) 
+36m^2\left(A_mA_n+4g_{mn}A^2\right) \cr
&+12mA_{(m}\partial_{n)}\sigma + 30mg_{mn}A^m\partial_m\sigma
+144m^2g_{mn}e^{-2\sigma} 
\ , \cr
D^nF_{mn} = &18mA_nF_m^{\ n} + 72m^2e^{-2\sigma}A_m 
- 24me^{-2\sigma}\partial_m\sigma 
\ ,\cr
6D^2\sigma - 8 (D\sigma)^2 = &R +{1\over4}F^2 + 360m^2e^{-2\sigma}
+288m^2A^2
+96mA^m\partial_m\sigma\cr 
& - 36mD^mA_m \ , \cr}
\eqn\eqom
$$
where $F_{mn}=\partial_mA_n-\partial_nA_m$.
In the above equations all curvatures and covariant 
derivatives are with respect to the ten dimensional Levi-Civita 
connection $\omega$. Clearly if $m=0$ these equations reduce to the
standard equations of type IIA supergravity [\CWHMGP]. Furthermore one
can check that
these equations are self-consistent in the sense the the intergrability
condition for the second equation in \eqom\ is implied by the other two
equations.

The form of the equations of motion is peculiar and it is not to difficult to 
see that they cannot be obtained from
a Lorentz invariant Lagrangian. Therefore the theory cannot be turned
into Romans supergravity by any field redefinition.
It is also clear from \eqom\ that the $U(1)$ symmetry of the gauge field $A_m$
has been affected  by the mass terms. However, a modified non-compact symmetry  
still exists and the fields equations are invariant
under the transformations
$$\eqalign{
A_m &\rightarrow A_m - \partial_m\lambda \ ,\cr
\sigma &\rightarrow \sigma -3m\lambda\ ,\cr
g_{mn} &\rightarrow e^{-6m\lambda}g_{mn}\ .\cr}
\eqn\symone
$$
Thus for $m\ne 0$ the dilaton $\sigma$ is eaten by the vector field $A_m$
which has become massive. If we had not discarded the antisymmetric tensor 
field in the compactification then 
there would also be a 2-form  gauge symmetry because $\hB_3$ 
always appears in the equations of motion  through its field strength
$\hH_4 ={\hat D}\hB_3$. Hence $\hB_3$ is defined only up to transformations
$\delta\hB_3 = {\hat D}\Lambda_2$. In ten dimensions where we may write 
$\hB_3 = C_3 + B_2\wedge dy$ this symmetry becomes
$$\eqalign{
C_3 &\rightarrow C_3 + d\Lambda_2 \ ,\cr
B_2 &\rightarrow B_2 + d\Lambda_1 + 6m\Lambda_2 \ .\cr}
\eqn\symmtwo
$$
Therefore if $m\ne 0$ the 3-form gauge field eats the 2-form and also becomes
massive, analogously to the  way that we arrived at massive electrodynamics in 
the
previous section.  

\chapter{A Euclidean Eightbrane}

In this section we will look for solutions to the field equations which
preserve half the supersymmetry. If we look for purely bosonic solutions
then we may just 
consider the supersymmetry transformation (written in eleven dimensional
form for simplicity)
$$
\delta{\hat \psi}_a  = \partial_a\epsilon + 
{1\over4}{\hat \omega}_a^{\ \ub\uc}\Gamma_{\ub\uc}
- m\he_{a}^{\ \ub}\he^{y\uc}\Gamma_{\ub \uc}\epsilon + 2m\delta_a^y\epsilon
+ {1\over36}\left(\epsilon\Gamma^{bcd}{\hat H}_{abcd}
+{1\over8}\epsilon\Gamma_{abcde}{\hat H}^{bcde}\right)\ .
\eqn\susy
$$
Note that the fourth term appearing in  $\delta{\hat \psi}_a$ does not
come with gamma matrix as it does in standard cosmological  supergravities
such as Romans. 
Let us look for an eightbrane solution with nine dimensional
Poincare symmetry and a single spacelike transverse space with the
fields ${\hat H_{abcd}}$ set to zero.  This immediately runs into
problems since the second and third terms appearing 
in $\delta{\hat \psi}_a$ contain
$\Gamma_{\ub\uc}$ which, for spacelike indices, has imaginary 
eigenvalues while the fourth term always
has real eigenvalues. This 
mis-match comes about because we have identified the compact 
$U(1)$ symmetry of
diffeomophisms of the circle with noncompact scale transformations. 
At the level
of the field equations there is no difficulty to doing this, since they 
have the
same Lie algebra, but it is reflected by the reality of $m$ in the 
supersymmetry
transformations \susy.
Thus there
is no eightbrane BPS soliton. However, this suggests that we look for BPS
solutions which are time dependent. 

We therefore will start with the ansatz
$$\eqalign{
ds^2_{10} &= -e^{2g}dt^2 + e^{2f}d{\bf x}\cdot d{\bf x}\ ,\cr 
F_{mn}& = 0\ ,\cr}
\eqn\an
$$
with $f,g,\sigma$ functions of time $t$ only. Demanding that this solve the 
field equations gives
$$\eqalign{
(\dot f - \dot \sigma)e^{\sigma -g} &= \pm 2m\ , \cr
A_m &= {1\over 3m}\partial_m \sigma\ ,}
\eqn\field
$$
where a dot denotes differentiation with respect to $t$. 
If we now ask that the ansatz \an\ preserves some supersymmetries we find that
$$\eqalign{
{\dot f}e^{\sigma-g}&=\pm 14m\ ,\cr
{\dot \sigma}e^{\sigma-g}&=\pm 12m\ ,\cr}
\eqn\susyeq
$$
and the remaining supersymmetries are generated by
$$\eqalign{
\epsilon &= e^{-3\sigma/4}\epsilon_0\ , \cr
\epsilon_{0}&=\mp\Gamma_{\uo}\Gamma_{\uy}\epsilon_0\ . \cr}
\eqn\susyres
$$
Clearly \susyeq\ implies that the field equation \field\ is solved. 
Note that by
a redefinition of $t$ 
we are free to choose $g$ to be any non-singular function. 
Let content ourselves here with $g=\sigma$. We then
find the solutions
$$\eqalign{
{\sigma} &= {\sigma_0} \pm 12mt \ ,\cr
f &= f_0 \pm14mt \ .\cr}
\eqn\solution
$$
This solution is similar to an eightbrane solution except that it has nine 
dimensional
Euclidean symmetry and the transverse space is timelike. At $t=0$ this solution
is simply flat ${\bf R}^9$. However, as time passes the radius of the compact
direction increases or decreases depending on the choice of sign in \solution.
For the plus sign the spacetime decompactifies back up to eleven dimensions
but for the minus sign it compactifies even further. We may lift the 
``Euclidean eightbrane'' solution to eleven dimensions
where it takes the form
$$
ds^2_{11} = e^{4\sigma/3}(-dt^2+dy'^2) + e^{5\sigma/3}d{\bf x}\cdot d{\bf x}\ ,
\eqn\eleven
$$
where $y' = y \pm 4t$. 

\chapter{Conclusions}

In this paper we have investigated the generation of mass in
supergravity
through compactification  on a circle and we have found a
massive supergravity in ten dimensions which is not that obtained by
Romans. The resulting theory has a number of unusual features, one of which 
is that it has no action. In addition the compactification ansatz used in this 
construction relates the internal compact symmetry of the circle to the
non-compact dilation symmetry. This is only required at the Lie algebra
level, where the two algebras are isomorphic, but it appears to be responsible
for the wrong sign for $m^2$ and
the time dependence of the supergravity BPS solution.

Can we provide the supergravity obtained here
with a natural interpretation?
One interpretation of $p$-branes is that they interpolate between different
supersymmetric vacua of a theory [\GT]. Even though the BPS state constructed
here is not a $p$-brane we may try to give it a similar interpretation. As 
$t\rightarrow -\infty$ (we assume the $-$ sign without loss of generality) 
the radius of the eleventh dimension blows up and
they theory decompactifies. However, at late times
the radius shrinks to zero. Thus this solution appears to be interpolating
between uncompactified M theory and weakly coupled type IIA string theory.
In the flat space  example in section three the vector field
in the lower dimension was tachyonic, indicating that the ground state is
unstable. The fact that the solution \an\ is time dependent suggests
that the ten dimensional theory is in fact unstable against decompactification
back up to eleven dimensions. Or, conversely, in a  time reversed picture
the theory is unstable against further compactification.

\refout

\end